\documentclass[preprint]{revtex4-1}
\usepackage{graphicx}
\def\be{\begin{equation}}\def\ee#1{\label{#1}\end{equation}}
\def\ba{\begin{array}}	\def\ea{\end{array}}
\def\bea{\begin{eqnarray}}	\def\eea{\end{eqnarray}}
\def\mc{\mathcal}  \def\pa{\partial}
  \def\ci{\cite}  \def\s{{\mbox{\scriptsize s}}}
\def\ra{\rightarrow}		
	
\def\N{{\mbox{\scriptsize N}}}	\def\S{{\mbox{\scriptsize S}}}
		\def\L{{\mbox{\scriptsize L}}}
\def\TOT{{\mbox{\scriptsize TOT}}} \def\S{{\mbox{\scriptsize S}}}
\def\QCD{{\mbox{\scriptsize QCD}}} 
 
\def\Re{{\rm Re}}   \def\Im{{\rm Im}} 

\def\Title#1{\begin{center} {\Large #1 } \end{center}}
\def\Author#1{\begin{center}{\sc #1} \end{center}}
\def\Address#1{\begin{center}{\it #1} \end{center}}

\newcommand\pubnumber{MAXLA-4/20, ``Constraint-RQM''}
\newcommand\pubdate{\today}
\newcommand\pubblock{\rightline{\begin{tabular}{l} \pubnumber\\
\pubdate \end{tabular}}}

\newenvironment{Presented}
{\begin{quotation} 
\begin{center} 
 PRESENTED AT\end{center}  
\begin{center}\begin{large}}{\end{large}
\end{center} 
\end{quotation}}

\begin{document}

\begin{titlepage}
\pubblock

\vfill
\Title{ Constraint Dynamics and RQM Bound States} 
\vspace{1mm}
\Author{Mikhail N. Sergeenko}
\Address{Fr. Skaryna Gomel State University, 
BY-246019, Gomel, Belarus\\
{\rm mnsergey@tut.by}}
\vspace{5mm}

\centerline{\bf Abstract}
\vspace{3mm}
 Flavored mesons containing quarks of unequal masses are studied. 
 The appropriate tool is the Bethe-Salpeter formalism, but its 
inherent complexity leads to series of difficulties mostly related 
to the central role played in it by the relative time or energy. 
 We consider bound states in the spirit of ``Constraint Relativistic 
Quantum Mechanics (RQM)''. 
 Interaction of quarks is described by the funnel-type potential 
with the distant dependent strong coupling, $\alpha_\s(r)$. 
 Relativistic bound-state problem is formulated with the use of 
symmetries, energy-momentum conservation laws in Minkowskiy space. 
 Relativistic two-body wave equation with position dependent particle 
masses is derived and used to describe the flavored mesons. 
 Free particle hypothesis for the bound state is developed: 
quark and antiquark move as free particles in of the bound system. 
 Solution of the equation for the system in the form of a~standing 
wave is given. 
 Interpolating complex-mass formula for two exact asymptotic eigenmass 
expressions is obtained. 
 Mass spectra for some leading-state flavored mesons are calculated. 
\vskip 5mm

\vspace{3mm}
\begin{Presented}
NONLINEAR PHENOMENA IN COMPLEX SYSTEMS \\
XXVII International Seminar\\
Chaos, Fractals, Phase Transitions, Self-organization \\
May 19--22, 2020, Minsk, Belarus 
\end{Presented}
\vspace{5mm}
\end{titlepage}

\section{Introduction}\label{intro}
 Most numerous of hadrons summarized in the Particle Data Group 
(PDG) tables~\ci{RPP2018} are mesons. 
 Simplest of them are quarkonia, quark-antiquark ($q\bar q$ and 
$Q\bar Q$) bound states containing quarks of equal masses. 
 It is believed that physics of light and heavy mesons is different, 
but this is true only in asymptotic limits of large and small 
distances. 
 In case of heavy-light $Q\bar q$ mesons situation is much more 
complicated.  
 Most mesons listed in the PDG being unstable and are resonances, 
exited quark-antiquark states. 
 There are great amount and variety of experimental data and 
the different approaches used to extract the properties of 
the mesons~\ci{KlemZaits07,NussLamp02,LiMaLi04}. 

 The strict description of mesons as quark-antiquark bound states 
in a way fully consistent with all requirements imposed by special 
relativity and Quantum Mechanics (QM) is one of the great challenges 
in theoretical elementary particle physics. 
 Such description can be done within the framework of Quantum Field 
Theory (QFT). 
 The appropriate tool to achieve this goal is the Bethe-Salpeter (BS) 
formalism~\ci{SalBeth51a}. 
 However, attempts to apply the BS formalism to relativistic 
bound-state problems lead to a number of various difficulties. 
 The usual practice consists in eliminating the relative time variable 
(3D reduction). 
 The 3D reduction of the two-fermion BS equation has been performed 
in many 
works~\ci{Salpet52,LogTav63,BlanSug66,Todor76,Sazd88,JallSaz97}; 
all these methods are theoretically equivalent. 
 There were suggested noncovariant instantaneous truncations of 
the BS equation~\ci{Nakan69,LuchSho16}. 
 The most well-known of the R2B equations is the one proposed by 
Salpeter~\ci{Salpet52}. 
 The equivalence of simplified equations with the original BS equation 
can be proved exactly~\ci{BijtBro97}. 

 In QED, the Coulomb gauge is the most convenient for treating the R2B 
problem, since it allows the optimal expansion of the BS equation 
around the NR theory~\ci{BodYen78,Naka69,Murota88}. 
 The main disadvantage of the Coulomb gauge is its noncovariant nature. 
 Constraint theory leads to a manifestly covariant 3D description 
of relativistic two-body (R2B) 
systems~\ci{Todor76,Komar78,Leut78,CratVanA83,LongLus86,CratVanA04} 
and has opened a new perspective. 

 It was shown~\ci{Todor76,JallSaz97} that the expansion of 
the BS equation around the constraint theory wave equations in 
the Feynman gauge (as well as for scalar interactions) is free of 
the above mentioned diseases of covariant gauges and allows a systematic 
study of infrared leading effects of multiphoton exchange diagrams; 
the latter can then be represented in three-dimensional (3D) x-space as 
local potentials. 
 Summing the series of these leading terms one obtains a local potential 
in compact form~\ci{JallSaz97}, which is well suited for a continuation 
to the {\it strong coupling domain} (QCD) of the theory or for 
a generalization to other effective interactions.

 In this work, we study $Q\bar q$ mesons and their excitations 
(resonances) as R2B systems from unified point of view in the framework 
of the RQM~\ci{GreinerRQM00,Dirac49}. 
 Difficulties encountered here are related to 
1)~two-particle relativistic equation of motion and 
2)~absence of a strict definition of the potential in relativistic 
theory. 
  We begin our consideration of the R2B problem with relativistic 
classical mechanics. 
 Relativistic bound state problem is formulated with the use of 
symmetries, energy-momentum conservation laws in Minkowskiy space. 
 The potential of interaction is treated as the Lorentz-scalar function 
of the spatial variable $r$, the distance between particles. 
 The concept of position dependent particle mass is developed. 
 Using the correspondence principle, we deduce, from the R2B classic 
equation, the two-particle wave equation. 
 The free particle hypothesis for the bound state is developed: particles 
inside the system move as free ones. 
 Complex eigenmasses for the bound system are obtained. 
 The relative motion of quarks in eigen states is described (in 
the physical region) by the standing wave of the form 
$C_n\sin(k_nx+\delta_n)$ for each spatial degree of freedom. 
 To verify the model, the R2B wave equation with position dependent 
quark masses is used to describe the flavored $Q\bar q$ mesons.

\section{Bound states in Constraint Theory}\label{ConstrTheor}
 The constraint theory first successfully yielded a~covariant yet 
canonical formulation of the R2B problem for two interacting 
spinless classical particles~\ci{LongLus86}. 
 The manifestly covariant formalism with constraints in the R2B 
problem leads to a Poincar'e invariant description of 
the dynamics of the system~\ci{CratVanA04,Sazd88}. 
 The potentials that appear in the corresponding wave equations 
can be calculated in terms of the kernel of the BS equation, 
therefore allow one to deal with QFT problems. 

 The R2B state of two scalar particles can be described by two 
independent wave equations, which are generalizations of 
the individual Klein-Gordon equation of each particle, including 
the mutual interaction potential~\ci{JallSaz97}. 
 The compatibility condition of the two equations imposes certain 
restrictions on the structure of the potential and leads, in 
a covariant form, to an eli\-mi\-nation of the relative energy variable. 
 This results in the manifestly covariant, 3D eigenvalue equation 
that describes the relative motion of the two particles~\ci{JallSaz97}. 
 This equation is very si\-mi\-lar in form to the Schr\"odinger 
(or Klein-Gordon) equation: it is a second order differential equation 
 in the three spacelike coordinates and therefore the usual techniques 
of NR QM are applicable to it. 

 For two fermions, the system is described by two independent Dirac 
type equations~\ci{CratVanA04}. 
 In this case, the compatibility condition imposes restrictions on 
the stucture of the potentials and eliminates the relative energy 
variable; however, because of the Dirac matrices, the reduction to 
a final eigenvalue equation is not straightforward. 
 The reduction process is rather complicate and depends on the way 
of eliminating the components of the spinor wave function in terms 
of one of them. 
 Up to now, no single Pauli-Schr\"odinger-type equation was obtained 
from this procedure. 

 The principal difficulty in the treatment of the BS equation 
(\ref{BSeq}) comes from the existence of unphysical relative time and 
energy variables. 
 In applications to both quantum electrodynamics (QED) and quantum 
chromodynamics (QCD), as well as NR reduction, some simplified 
equations are usually used. 
 Such equations are usually obtained using certain restrictions or 
constraints.

\subsection{The spinless Salpeter equation}\label{SSequ}
 The manifestly covariant BS equation obtained directly from QFT 
governs all the bound states and the scattering. 
 However, attempts to apply the BS formalism to real relativistic 
R2B problems lead to a number of various difficulties. 
 These are the impossibility to determine the BS interaction kernel 
beyond the tight limits of perturbation theory, appearance of 
abnormal solutions that are difficult to interpret in the framework 
of quantum physics. 
 Two body BS equation~\ci{BethSal08,SalBeth51,LuchSho16} for 
spin-zero bound states is
\be
G_0^{-1}\Psi \equiv\left(p_1^2 + m_1^2\right)
\left(p_2^2 + m_2^2\right)\Psi = K\Psi,
\ee{BSeq}
where $G_0=G_{0,1}G_{0,2}$ is free propagator of particles. 
 The irreducible BS kernel $K$ would in general contain charge 
renormalization, vacuum polarization graphs and could contain 
self-energy terms transferred from the inverse propogators. 
 The kernel $K$ is obtained from the off-mass-shell scattering 
amplitude which is defined by the equation $T = K + KG_0T$. 
 Recent work with static models has indicated, that abnormal 
solutions disappear if one includes all ladder and cross ladder 
diagrams~\ci{JallSaz97}. 
 This supports Wick's conjecture on defects of ladder 
approximations. 

 Numerous 3D quasipotential reductions of the BS equation had 
been proposed. 
 The most well-known of the resulting bound-state equations is 
the one proposed by Salpeter~\ci{Salpet52}. 
 The Salpeter equation is historically first 3D reduction of 
the BS equation. 
 This is a noncovariant instantaneous truncations of the BS 
equation (\ref{BSeq}). 
 The general coordinate-space relativistic spinless Salpeter (SS) 
equation for R2B system in the c.m. rest frame is ($\hbar=c=1$)
\be
\left[\sqrt{(-i\vec\nabla)^2 + m_1^2} + \sqrt{(-i\vec\nabla)^2 +
m_2^2} + V(r)\right]\psi(\vec{r}) = {\sf M}\psi(\vec{r}), 
\ee{SSeq}
where $V(r)$ is the potential of interaction and ${\sf M}$ is 
the bound-system mass (the c.m. rest energy ${E^*=\sf M}=w$). 
 However, as in the case of the BS equation, it is a problem 
to find the analytic solution of the equation (\ref{SSeq}). 
 The problem originates from two square root operators which cause 
a serious difficulties.
 It can not be reduced to the second-order differential equation 
of the Shr\"odinger type~\ci{Todor76}. 

 There exist many other approaches to bound-state problem. 
 One of the promising among them is the Regge method in hadron 
physics~\ci{Collin77}. 
 All hadrons and their resonances in this approach are associated 
with Regge poles which move in the complex angular momentum $J$~plane. 
 Moving poles are described by the Regge trajectories, $\alpha(s)$, 
which are the functions of the invariant squared mass $s=w^2=(E^*)^2$ 
(Mandelstam's variable). 
 Hadrons and resonances populate their Regge trajectories which 
contain all the dynamics of the strong interaction in bound state 
and scattering regions. 

 Mesons have been studied in a soft-wall holographic approach 
AdS/CFT~\ci{Lyubo10} using the correspondence of string theory in 
Anti-de Sitter space and conformal field theory in physical space-time. 
 It is analogous to the Schr\"odinger theory for atomic physics 
and  provides a precise mapping of the string modes $\Phi(z)$ in 
the AdS fifth dimension $z$ to the hadron light-front wave functions 
in physical space-time. 

 Reductions of the BS equation can be obtained from iterating this 
equation around a 3D Lorentz-invariant hypersurface in relative 
momentum ($p$) space. 
 This leads to invariant 3D wave equations for relative motion. 
 The resultant 3D wave equation is not unique, but depends on 
the nature of the 3D hypersurface.

\subsection{Todorov's quasipotential equation}\label{Todequ}
 Valuable are methods which provide either exact or approximate 
analytic solutions for various forms of differential equations. 
 They may be remedied in 3D reductions of the BS equation. 
 In most cases the analytic solution can be found if original 
equation is reduced to the Schr\"odinger-type wave equation. 
 One can choose Todorov's quasipotential equation~\ci{Todor76} 
which has the Schr\"odinger-like form
\be
\left[\hat p^2+\Phi(x_1-x_2)\right]\psi = \kappa_w^2\psi,
\ee{TodorEq}
where the quasipotential $\Phi$ is related to the scattering amplitude, 
3D hyperfine restriction on the relative momentum $p$ is defined by 
$p\cdot P\psi=0$, $\hat{p}\psi=(0,\hat{\mathbf p})\psi=0$, $P=p_1+p_2$. 
 The effective eigenvalue in (\ref{TodorEq}) is 
\be
\kappa_w^2 
= \frac 1{4w^2}\left(w^2 - m_-^2\right)\left(w^2 - m_+^2\right),
\ee{EffMom2}
with $w=\sqrt{P^2}=E^*$ the c.m. invariant energy, $m_-=m_1-m_2$, 
$m_+=m_1+m_2$. 
 The forces $\Phi$ to depend on $x_1 - x_2$ only through the transverse 
component, $x_\perp^\mu=(0,\mathbf r)$. 
 Thus, in the c.m. rest frame, the hypersurface restriction 
$p\cdot P\psi=0$ not only eliminates the relative energy but implies 
that the relative time does not appear. 

 In Eikonal approximation for ladder, cross ladder, and constraint 
diagrams to bound states applied through all orders, it gives for 
scalar ($S$) and vector ($V$) exchanges the quasipotentials 
\be
\Phi_\s = 2m_w S + S^2,\quad \Phi_V = 2\epsilon_w V - V^2. 
\ee{ScaVecPot}
 The kinematical variables (the reduced mass $m_w$ and energy 
$\epsilon_w$ for the fictitious particle of relative motion), 
\be
m_w = \frac{m_1 m_2}w,\quad \epsilon_w =\frac{w^2-m_1^2-m_2^2}{2w},
\ee{RelMaEn}
satisfy the Einstein's relation
\be
\kappa_w^2 = \epsilon_w^2 - m_w^2.
\ee{EinRel}
 The effects of ladder and cross ladder diagrams thus embedded in 
their c.m. energy dependencies. 
 The resultant 3D wave equation (\ref{TodorEq}) is not unique, 
but depends on the nature of the 3D hypersurface.

\subsection{The two-body Dirac equations}\label{Todequ}
 For two fermions, the system can be described by two independent 
Dirac type equations~\ci{CratVanA83,LongLus86,CratVanA04}. 
 In this case, the compatibility condition imposes restrictions on 
the stucture of the potentials and eliminates the relative energy 
variable; however, because of the Dirac matrices, the reduction to 
a final eigenvalue equation is not straightforward. 
 The reduction process is rather complicate and depends on the way 
of eliminating the components of the spinor wave function in terms 
of one of them. 
 Up to now, no single Pauli-Schr\"odinger type equation was obtained 
from this procedure. 

 The R2B Dirac equations of Constraint Dynamics have dual origins. 
 On the one hand they arise as one of the many quasipotential 
reductions of the BS equation. 
 On the other they arise independently from the development of 
a consistent covariant approach to the R2B problem in the framework 
of RCM independent of QFT~\ci{CratVanA83}. 

 The R2B Dirac equations of constraint dynamics provide a covariant 
3D truncation of the BS equation. 
 It was shown~\ci{SazdProc92,Suzd87} that the BS equation can be 
algebraically transformed into two independent equations, which for 
spinless particles are 
\bea
\left(\mc{H}_{0,1} +\mc{H}_{0,2} +2\Phi_w\right)\Psi(x_1,\,x_2) = 0,
\label{H1pH2}\\
\left(\mc{H}_{0,1}-\mc{H}_{0,2}\right)\Psi(x_1,\,x_2)
=2(p\cdot P)\Psi(x_1,\,x_2)=0,\label{H1mH2}
\eea
where $\mc{H}_{0,i}= p_i^2 + m_i^2$, $P=p_1+p_2$ is the total momentum, 
$p=\eta_2p_1 -\eta_1p_2$ is the relative momentum, $w$ is the invariant 
total c.m. energy with $P^2=-w^2$, $\hat{P}^\mu=P^\mu/w$ is a time-like 
unit vector ($\hat{P}^2=-1$) in the direction of the total momentum. 
 The $\eta_i$ are {\em chosen} so that the relative coordinate 
$x=x_1 - x_2$ and $p$ are canonically conjugate, i.e. $\eta_1+\eta_2=1$. 
 The first equatlity (\ref{H1pH2}) is a covariant 3D eigenvalue equation. 
 The second equation (\ref{H1mH2}) overcomes the difficulty of treating 
the relative time in the c.m. system by setting an invariant condition 
on the relative momentum $(p\cdot P)\Psi(x_1,\,x_2)=0$, that implies 
$p^\mu\Psi=p_\perp^\mu\Psi$.

\section{ The interaction potential}\label{ThePoten}
  It is well known that the potential as a~function in 3D-space 
is defined by the pro\-pa\-ga\-tor $D(\mathbf{q}^{\,2})$ (Green 
function) of the virtual particle as a carrier of interaction, where 
$\mathbf{q}=\mathbf{p}_1-\mathbf{p}_2$ is the transfered momentum. 
 In case of the Coulomb potential the propagator is 
$D(\mathbf{q}^{\,2})=-1/\mathbf{q}^{\,2}$; the Fourier transform 
of $4\pi\alpha D(\mathbf{q}^{\,2}$) gives the Coulomb potential, 
$V(r)=-\alpha/r$. 
 The relative momentum $\mathbf{q}$ is conjugate to the relative vector 
$\mathbf{r}=\mathbf{r}_1-\mathbf{r}_2$, therefore, one can accept 
that $V(\mathbf{r}_1,\,\mathbf{r}_2)=V(\mathbf{r})$~\ci{LuchSho99}. 
 If the potential is spherically symmetric, one can write  
$V(\mathbf{r})=>V(r)$, where $r=|\mathbf{r}|$.  
 Thus, the system's relative time $\tau=t_1-t_2=0$ 
(instantaneous interaction). 

 The NR QM shows very good results in describing bound states; 
this is partly because the potential is NR concept. 
 In relativistic mechanics one faces with different kind of 
speculations around the potential, because of absence of a strict 
definition of the potential in this theory. 
 In NR formulation, the hydrogen ($H$) atom is described by 
the Schr\"odinger equation and is usually considered as an electron 
moving in the external field generated by the proton static 
electric field given by the Coulomb potential.  
 In relativistic case, the binding energy of an electron in a~static 
Coulomb field (the external electric field of a point nucleus of 
charge $Ze$ with infinite mass) is determined predominantly by 
the Dirac eigenvalue~\ci{MohrTayl}. 
 The spectroscopic data are usually analyzed with the use of 
the Sommerfeld's fine-structure formula~\ci{Bohm79}, 

 One should note that, in these calculations the $S$ states start 
to be destroyed above $Z=137$, and that the $P$ states being 
destroyed above $Z=274$. 
 Similar situation we observe from the result of the Klein-Gordon 
wave equation, which predicts $S$ states being destroyed above $Z=68$ 
and $P$ states destroyed above $Z=82$. 
 Besides, the radial $S$-wave function $R(r)$ diverges as $r\ra 0$. 
 These problems are general for all Lorentz-vector potentials which 
have been used in these calculations~\ci{Huang01,Bhadur95}. 
 In general, there are two different relativistic versions: 
the potential is considered either as the zero component of 
a~four-vector, a~Lorentz-scalar or their mixture~\ci{SahuAll89}; 
its nature is a~serious problem of relativistic potential 
models~\ci{Sucher95}. 

 This problem is very important in hadron physics where, for 
the vector-like confining potential, there are no normalizable 
solutions~\ci{Sucher95,SemayCeu93}. 
 There are normalizable solutions for scalar-like potentials, 
but not for vector-like. 
 This issue was investigated in~\ci{MyZPhC94,Huang01};  
it was shown that the effective interaction has to be Lorentz-scalar 
in order to confine quarks and gluons. 
 The relativistic correction for the case of the Lorentz-vector 
potential is different from that for the case of the Lorentz-scalar 
potential~\ci{MyMPLA97}. 

 Quarkonia among all mesons are simplest as quark-antiquark bound 
states. 
 The quarkonium universal mass formula and ``saturating'' Regge 
trajectories were derived in~\ci{MyZPhC94} and in~\ci{MyEPJC12,MyEPL10} 
applied for gluonia (glueballs). 
 The mass formula was obtained by interpolating between NR heavy 
$Q\bar Q$ quark system and ultra-relativistic limiting case of light 
$q\bar q$ mesons for the Cornell potential~\ci{BaliLat01,EichGMR08}, 

\be 
V(r)=V_\S(r)+V_\L(r)\equiv-\frac 43\frac{\alpha_\s}r +\sigma r.
\ee{CornPot} 
 The short-range Coulomb-type term $V_\S(r)$, originating from 
one-gluon exchange, dominates for heavy mesons and the linear one 
$V_\L(r)$, which models the string tension, dominates for light mesons. 
 Parameters $\alpha_\s$ and $\sigma$ are directly related to basic 
physical quantities of mesons. 

 Separate consideration of two asymptotic components $V_\S(r)$ and 
$V_\L(r)$ of the potential (\ref{CornPot}) for quarkonia results in 
the complex-mass expression for resonances, which in the 
center-of-momentum (c.m.) frame is ($\hbar=c=1$)~\ci{MyAHEP13,MyNPCS14}: 
\be 
\mc{M}_\N^2 = 4\left[\left(\sqrt{2\sigma\tilde{N}}
+\frac{i\tilde\alpha m}N \right)^2
+\left(m-i\sqrt{2\tilde\alpha\sigma}\right)^2\right],
\ee{CompE2n}
where $\tilde\alpha=\frac 43\alpha_\s$, $\tilde{N}=N+(k+\frac12)$, 
$N=k+l+1$, $k$ is radial and $l$ is orbital quantum numbers; it has 
the form of the squared energy 
$\mc{M}_\N^2=4\left[(\pi_\N)^2+\mu^2\right]$ of two free relativistic 
particles with the quarks' complex momenta $\pi_\N$ and masses 
$\mu$. 
 This formula allows to calculate in a~unified way the centered 
masses and total widths of heavy and light quarkonia.  
 In our method the energy, momentum and quark masses are {\it complex}. 

 The Cornell potential (\ref{CornPot}) is a~special in hadron physics 
and results in the complex energy and mass eigenvalues. 
 As known, operators in ordinary QM are Hermitian and the corresponding eigenvalues are real. 
 It is possible to extend the Hamiltonian in QM into the complex domain 
while still retaining the fundamental properties of a~quantum theory. 
 One of such approaches is {\it complex} QM~\ci{BendBH}. 
 The complex-scaled method is the extension of theorems and principles 
proved in QM for Hermitian operators to non-Hermitian operators.

\subsection{ Modification of the Cornell potential}\label{ThePoten}
 The Cornell potential (\ref{CornPot}) is fixed by the two free 
parameters, $\alpha_\s$ and $\sigma$. 
 However, the strong coupling $\alpha_\s$ in QCD is a~function 
$\alpha_\s(Q^2)$ of virtuality $Q^2$ or $\alpha_\s(r)$ in configuration 
space. 
 The potential can be modified by introducing the 
$\alpha_\s(r)$-dependence, which is unknown. 
 A~possible modification of $\alpha_\s(r)$ was introduced in~\ci{MyEPJC12}, 
\be
V_\QCD(r) = -\frac 43\frac{\alpha_\s(r)}r +\sigma r,\quad 
\alpha_\s(r)=
 \frac 1{b_0\ln[1/(\Lambda_\QCD r)^2+(2\mu_g/\Lambda_\QCD)^2]}, 
\ee{VmodCor}
where $b_0=(33-2n_f)/12\pi$, $n_f$ is number of flavors, 
$\mu_g=\mu(Q^2)$ --- gluon mass at $Q^2=0$, $\Lambda_\QCD$ is the QCD 
scale parameter. 
 The running coupling $\alpha_\s(r)$ in (\ref{VmodCor}) is frozen at  
$r\ra\infty$, $\alpha_\infty=\frac 12[b_0\ln(2\mu_g/\Lambda_\QCD)]^{-1}$, 
and is in agreement with the asymptotic freedom properties, i.\,e., 
$\alpha_\s(r\ra 0)\ra 0$. 

 A more complicate case are flavored heavy-light $Q\bar q$ mesons. 
 A simplest example of heavy-light two-body system is the $H$ atom, 
comprising only a~proton and an~electron which are stable particles. 
 This simplicity means its properties can be calculated theoretically 
with impressive accuracy~\ci{MyRelHx19}. 
 The~spherically symmetric Coulomb potential, with interaction strength 
parametrized by dimensionless coupling (``fine structure'') constant 
$\alpha$, is of particular importance in many realms of physics.  
 The $H$ atom can be used as a tool for testing any relativistic 
two-body theory, because latest measurements for transition 
frequencies have been determined with a highest 
precision~\ci{MohrTayl}.

\section{ The R2B wave equation and its solution}\label{SolQCEq}
 Standard relativistic approaches for R2B systems run into serious 
difficulties in solving known relativistic wave equations. 
 The formulation of RQM differs from NR QM by the replacement of 
invariance under Galilean transformations with invariance under 
Poincar\`e transformations. 
 The RQM is also known in the literature as relativistic Hamiltonian 
dynamics or Poincar\`e-invariant QM with direct interaction~\ci{Dirac49}. 
 There are three equivalent forms in the RQM called ``instant'', 
``point'', and ``light-front'' forms. 

 The dynamics of many-particle system in the RQM is specified 
by expressing ten generators of the Poincar\`e group, 
$\hat M_{\mu\nu}$ and $\hat W_\mu$, in terms of dynamical variables. 
 In the constructing generators for interacting systems it is customary 
to start with the generators of the corresponding non-interacting 
system; the interaction is added in the way that is consistent with 
Poincare algebra. 
 In the relativistic case it is necessary to add an interaction $V$ 
to more than one generator in order to satisfy the commutation 
relations of the Poincar\'e algebra. 

 The interaction of a~relativistic particle with the four-momentum 
$p_\mu$ moving in the external field $A_\mu(x)$ is introduced in 
QED according to the gauge invariance principle, 
$p_\mu\ra P_\mu=p_\mu-eA_\mu$. 
 The description in the ``point'' form of RQM implies that the mass 
operators $\hat M^{\mu\nu}$ are the same as for non-interacting 
particles, i.\,e., $\hat M^{\mu\nu}=M^{\mu\nu}$, and these 
interaction terms can be~presented only in the form of 
the four-momentum operators~$\hat W^\mu$~\ci{MyRQMAnd99}. 

 Consider the R2B problem in Relativistic Classic Theory (RCM). 
 Two particles with four-momenta $p_1^\mu$, $p_2^\mu$ and 
the interaction field $W^\mu(q_1,\,q_2)$ together compose a~closed 
conservative system, which can be characterized by the 4-vector 
$\mc{P}^\mu$, 
\be 
\mc{P}^\mu = p_1^\mu + p_2^\mu + W^\mu(q_1,\,q_2),
\ee{Main4vec}
where the space-time coordinates $q_1^\mu$, $q_2^\mu$ and 
four-momenta $p_1^\mu$, $p_2^\mu$ are conjugate variables, 
$\mc{P}_\mu \mc{P}^\mu=\mathsf{M}^2$; here $\mathsf{M}$ is 
the system' invariant mass. 
 Underline, that no external field and each particle 
of the system can be considered as moving source of the interaction 
field; the interacting particles and the potential are a~unified 
system. 
 There are the following consequences of (\ref{Main4vec}) and 
they are key in our approach. 

 The four-vector (\ref{Main4vec}) describes {\it free motion} of 
the bound system and can be presented as two equations, 
\bea
E = \sqrt{\mathbf{p}_1^2+m_1^2}+\sqrt{\mathbf{p}_2^2+m_2^2}
+W_0(q_1,\,q_2)=\rm{const}, \quad \label{TwoEnr}\\ 
\mathbf{P}=\mathbf{p}_1+\mathbf{p}_2
+\mathbf{W}(q_1,\,q_2)=\rm{const}, \quad\label{TwoMom}
\eea
describing the energy and momentum conservation laws. 
 The energy (\ref{TwoEnr}) and total momentum (\ref{TwoMom}) 
of the system are the constants of motion. 
 By definition, for conservative systems, the integrals (\ref{TwoEnr}) 
and (\ref{TwoMom}) can not depend on time explicitly. 
 This means the interaction $W(q_1,\,q_2)$ should not depend on 
time, i.\,e., $W(q_1,\,q_2)=>V(\mathbf{r}_1,\,\mathbf{r}_2)$. 

 Equations (\ref{TwoEnr}) and (\ref{TwoMom}) in the c.m. frame are 
\bea 
\mathsf{M} = 
\sqrt{\mathbf{p}^2+m_1^2}+\sqrt{\mathbf{p}^2 +m_2^2}+\mathsf{V}(r),
\label{ClasE2B} \\
\mathbf{P}=\mathbf{p}_1 +\mathbf{p}_2 +
\mathbf{W}(\mathbf{r}_1,\,\mathbf{r}_2)=\mathbf{0}, \label{ClasM2B}
\eea
where $\mathbf{p}=\mathbf{p}_1 =-\mathbf{p}_2$ that follows from the 
equality $\mathbf{p}_1+\mathbf{p}_2=0$; this means that 
$\mathbf{W}(\mathbf{r}_1,\,\mathbf{r}_2)=0$. 
 The system's mass (\ref{ClasE2B}) in the c.m. frame is 
Lorentz-scalar. 
 In case of free particles ($\mathsf{V}=0$) the invariant mass 
$\mathsf{M}=\sqrt{\mathbf{p}^2+m_1^2}+\sqrt{\mathbf{p}^2+m_2^2}$ 
can be transformed for $\mathbf{p}^2$ as 
\be
\mathbf{p}^2
=\frac 1{4s}(s-m_-^2)(s-m_+^2)\equiv\mathsf{k}_\s^2, \quad\quad\quad  
s = \mathsf{M}^2. 
\ee{InvMom1}

 Equation (\ref{TwoEnr}) is the zeroth component of the four-vector 
(\ref{Main4vec}). 
 But, in the c.m. frame the mass (\ref{ClasE2B}) is Lorentz-scalar; 
and what about the potential $\mathsf{V}$? 
 Is it still Lorentz-vector? 
 To show that the potential is Lorentz-scalar, let us reconsider 
(\ref{ClasE2B}) as follows. 
 The relativistic total energy $\epsilon_i(\mathbf{p})$ ($i=1,\,2$) 
of particles in (\ref{ClasE2B}) given by 
$\epsilon_i^2(\mathbf{p})=\mathbf{p}^2+m_i^2$ can be represented as 
sum of the kinetic energy $\tau_i(\mathbf{p})$ and the particle rest 
mass $m_i$, i.\,e., $\epsilon_i(\mathbf{p})=\tau_i(\mathbf{p})+m_i$. 
 Then the system's total energy (invariant mass) (\ref{ClasE2B}) 
can be written in the form 
$\mathsf{M}=\sqrt{\mathbf{p}^2+\mathsf{m}_1^2(r)}
+\sqrt{\mathbf{p}^2+\mathsf{m}_2^2(r)}$, where 
$\mathsf{m}_{1,2}(r)=m_{1,2}+\frac 12\mathsf{V}(r)$ are 
the distance-dependent particle masses~\ci{MyNDA17} and (\ref{InvMom1}) 
with the use of $\mathsf{m}_1(r)$ and $\mathsf{m}_2(r)$ takes the form, 
\be
\mathbf{p}^2 = 
K_\s\left[s-(m_+ +\mathsf{V})^2\right]\equiv\mathsf{k}_\s^2-U(s,\,r),
\ee{InvMom2}
where $K_\s=(s-m_-^2)/4s$, $\mathsf{k}_\s^2$ is squared invariant 
momentum given by (\ref{InvMom1}) and 
$U(s,\,r) = K_\s\left[2m_+\mathsf{V} +\mathsf{V}^2\right]$ 
is the potential function. 
 The equation (\ref{InvMom2}) is the relativistic analogy of the NR 
expression $\mathbf{p}^2=2\mu[E-V(r)]\equiv\mathsf{k}_E^2-U(E,r)$. 

 The equality (\ref{InvMom2}) with the help of the fundamental 
correspondence principle gives the two-particle spinless wave 
equation,
\be 
\left[(-i\vec\nabla)^2 + U(s,\,r)\right]\psi(\mathbf{r})
= \mathsf{k}_\s^2\psi(\mathbf{r}). 
\ee{Rel2Eq}
 The equation (\ref{Rel2Eq}) can not be solved by known methods for 
the potential (\ref{VmodCor}). 
 Here we use the quasiclassical (QC) method and solve another wave 
equation~\ci{MyMPLA97,MyPRA96}.

\subsection{ Solution of the R2B wave equation}\label{SolQCEq}
 Solution of the Shr\"odinger-type's wave equation (\ref{Rel2Eq}) 
can be found by the QC method developed in~\ci{MyPRA96}. 
 In our method one solves the QC wave equation derivation of which 
is reduced to replacement of the operator $\vec{\nabla}^2$ in 
(\ref{Rel2Eq}) by the canonical operator $\Delta^c$ without 
the first derivatives, acting onto the state function 
$\Psi(\vec r)=\sqrt{{\rm det}\,g_{ij}}\psi(\vec r)$, where $g_{ij}$ 
is the metric tensor. 
 Thus, instead of (\ref{Rel2Eq}) one solves the QC equation, for 
the potential~(\ref{VmodCor}), 
\be 
\Biggl\{\frac{\pa^2}{\pa r^2}+\frac 1{r^2}\frac{\pa^2}{\pa\theta^2}
+\frac 1{r^2\sin^2\theta}\frac{\pa^2}{\pa\varphi^2}
+K_\s\biggl[s-\left(m_+ -\frac 43\frac{\alpha_\s(r)}r
+\sigma r\right)^2\biggr]\Biggr\}\Psi(\mathbf{r})=0.
\ee{Rel2Equa}
 This equation is separated. 
 Solution of the angular equation was obtained in~\ci{MyPRA96} by 
the QC method in the complex plane, that gives 
$\textrm{M}_l=(l+\frac 12)\hbar$, for the angular momentum eigenvalues. 
 These angular eigenmomenta are universal for all spherically symmetric 
potentials in relativistic and NR cases. 

 The radial problem has four turning points and cannot be solved by 
standard methods. 
 We consider the problem separately by the QC method for the short-range 
Coulomb term (heavy mesons) and the long-range linear term (light mesons). 
 The QC method reproduces the exact energy eigenvalues for all known 
solvable problems in QM~\ci{MyMPLA97,MyPRA96}. 
 The radial QC wave equation of (\ref{Rel2Equa}) for the Coulomb term 
has two turning points and the phase-space integral is found in 
the complex plane with the use of the residue theory and method of 
stereographic projection~\ci{MyPRA96,MyAHEP13} that gives 
\be
\mc{M}_\N^2|_C 
=\left(\sqrt{\epsilon_\N^2}\pm \sqrt{(\epsilon_\N^2)^*}\right)^2 
\equiv \Re\{\epsilon_\N^2\}\pm i\Im\{\epsilon_\N^2\},
\ee{W2Coul}
where $\epsilon_\N^2=m_+^2\left(1-v_\N^2\right)+2im_+ m_-v_\N$, 
$v_\N=\frac 23\alpha_\infty/N$, $N=k+l+1$. 

 Large distances in hadron physics are related to the problem of 
confinement. 
 The radial problem of (\ref{Rel2Equa}) for the linear term has four 
turning points, i.\,e., two cuts between these points. 
 The phase-space integral in this case is found by the same method 
of stereographic projection as above that results in the cubic 
equation~\ci{MyNDA17}: $s^3 + a_1s^2 + a_2s + a_3 = 0$, where 
$a_1=16\tilde\alpha_\infty\sigma-m_-^2$, 
$a_2=64\sigma^2\left(\tilde\alpha_\infty^2-\tilde N^2
-\tilde\alpha_\infty m_-^2/4\sigma\right)$, 
$a_3=-(8\tilde\alpha_\infty\sigma m_-)^2$, $\tilde N=N+k+\frac 12$, 
$\tilde\alpha_\infty=\frac 43\alpha_\infty$, 
$\alpha_\infty=\alpha_\s(r\ra\infty)$. 
 The first root $s_1(N)$ of this equation gives the physical solution 
(complex eigenmasses), $\mathsf{M}_1^2|_L=s_1(N)$, for the squared 
invariant mass. 

 Two exact asymptotic solutions obtained such a way are used to derive 
the interpolating mass formula. 
 The~interpolation procedure for these two solutions~\ci{MyZPhC94} 
is used to derive the mesons' complex-mass formula, 
\be
\mc{M}_\N^2 =\left(m_1+m_2\right)^2\left(1-v_\N^2\right)
\pm 2im_+m_-v_\N +\mathsf{M}_1^2|_L.
\ee{W2int}
 The real part of the square root of (\ref{W2int}) defines the~centered 
masses and its imaginary part defines the~total widths, 
$\Gamma_\N^\TOT=-2\,\Im\{\mathsf{M}_\N\}$, of mesons and 
resonances~\ci{MyAHEP13,MyNPCS14}. 

 In the QC method not only the total energy, but also momentum of 
a~particle-wave in bound state is the {\em constant of motion}. 
 Solution of the QC wave equation in the whole region is written 
in elementary functions as~\ci{MyPRA96}
\be
\mathsf{R}(r) = C_n\left\{\ba{lc} 
\frac 1{\sqrt 2}e^{|\mathsf{k}_n|r -\phi_1}, & r<r_1,\\
\cos(|\mathsf{k}_n|r -\phi_1 -\frac\pi 4), & r_1\le r\le r_2,\\
\frac{(-1)^n}{\sqrt 2}e^{-|\mathsf{k}_n|r +\phi_2}, & r>r_2,
\ea\right.
\ee{osol}
where $C_n=\sqrt{2|\mathsf{k}_n|/[\pi(n+\frac 12)+1]}$ is 
the normalization coefficient, $\mathsf{k}_n$ is the corresponding 
eigenmomentum found from solution of (\ref{Rel2Eq}), 
$\phi_1=-\pi(n+\frac 12)/2$ and $\phi_2=\pi(n+\frac 12)/2$ are 
the values of the phase-space integral at the turning points 
$r_1$ and $r_2$, respectively. 

 The free fit to the data~\ci{RPP2018} shows a~good agreement for 
the light and heavy $Q\bar q$ meson and their resonances. 
 To demonstrate efficiency of the model we calculate the leading-state 
masses ($S=1$) of the $\rho$ and $D^*$ meson resonances (see tables, 
where masses are in MeV). 
 Parameters of calculations are also in the tables. 
\begin{table}[ht]
\begin{center}
\caption{The masses of the $\rho^\pm$-mesons and resonances}
\label{rho_mes}
\begin{tabular}{lllll}
\hline\noalign{\smallskip}
\ \ Meson &~~~$J^{PC}$ &~~~$\ \ E_n^{ex}$ &~~~$\ \ E_n^{th}$&
~~~Parameters in (\ref{W2int})\\
\noalign{\smallskip}\hline\hline\noalign{\smallskip}
\ \ \ $\rho\ (1S)$&~~~$1^{--}$&~~~$\ \ 775.5$&~~~$\ \ 775.3$&
~~~~~$\alpha_\s=1.478$\\ 
\ \ \ $a_2(1P)$&~~~$2^{++}$&~~~$\ 1318.3$&~~~$\ 1317.9$&
~~~~~$\sigma=0.142$\,GeV$^2$\\ 
\ \ \ $\rho_3(1D)$&~~~$3^{--}$&~~~$\ 1688.8$&~~~$\ 1695.0$&
~~~~~$m_d=4.70$\,MeV\\ 
\ \ \ $a_4(1F)$&~~~$4^{++}$&~~~$\ 1996.6$&~~~$\ 2002.2$&
~~~~~$m_u=2.15$\,MeV\\ 
\ \ \ $\rho\ (1G)$&~~~$5^{--}$&~~~$\ 2330.0 $&~~~$\ 2268.3$&
~~~~~~~~~~~~\\
\ \ \ $\rho\ (2S)$&~~~$1^{--}$&~~~$\ 1720.0$&~~~$\ 1695.5$&~\\
\ \ \ $\rho\ (2P)$&~~~$2^{++}$&~~~$ ~ $&~~~$\  2002.2$&~\\
\ \ \ $\rho\ (2D)$&~~~$3^{--}$&~~~$ ~ $&~~~$\ 2268.3$&~\\
\noalign{\smallskip}\hline
\end{tabular}
\end{center}

\begin{center}
\caption{The masses of the $D^{\pm*}$-mesons and resonances}
\label{Dp_mes}
\begin{tabular}{lllll}
\hline\noalign{\smallskip}
\ \ Meson &~~~$J^{PC}$ &~~~$\ \ E_n^{ex}$ &~~~$\ \ E_n^{th}$&
~~~Parameters in (\ref{W2int})\\
\noalign{\smallskip}\hline\hline\noalign{\smallskip}
\ \ \ $D^*(1S)$&~~~$1^{--}$&~~~$\ 2010.3$&~~~$\ 2010.3$&
~~~~~$\alpha_\S=1.308$\\
\ \ \ $D_2^*(1P)$&~~~$2^{++}$&~~~$\ 2460.1$&~~~$\ 2432.0$&
~~~~~$\sigma=0.275$\,GeV$^2$\\ 
\ \ \ $D_3^*(1D)$&~~~$3^{--}$&~~~$\ ~ $&~~~$\ 2823.2$&
~~~~~$m_c=1026.9$\,MeV\\ 
\ \ \ $D_4^*(1F)$&~~~$4^{++}$&~~~$\ ~ $&~~~$\ 3176.9$&
~~~~~$m_d=4.7$\,MeV\\
\ \ \ $D_5^*(1G)$&~~~$5^{--}$&~~~$\ ~ $&~~~$\ 3499.3$&
~~~~~ \\
\ \ \ $D^*(2S)$&~~~$1^{--}$&~~~$\ ~ $&~~~$\ 2822.9$&~\\
\ \ \ $D^*(2P)$&~~~$2^{++}$&~~~$\ ~ $&~~~$\ 3176.8$&~\\
\ \ \ $D^*(2D)$&~~~$3^{--}$&~~~$\ ~ $&~~~$\ 3499.2$&~\\
\noalign{\smallskip}\hline
\end{tabular}
\end{center}
\end{table}
 The strong coupling constant, $\alpha_\s$, is given by (\ref{VmodCor}) 
and expressed via the gluon mass $m_g$ and the QCD scale parameter 
$\Lambda_\QCD$. 
 The gluon mass, $m_g=416$\,MeV, is the same for all types of mesons: 
$\rho^\pm$, $K$, $D$, $B$ mesons and also for glueballs~\ci{MyEPJC12}. 
 The QCD scale parameter $\Lambda_\QCD^\rho=638\,MeV$, for 
the $\rho$ mesons. 
 The relative error of the data description is $\epsilon^\rho=0.67\%$. 
 The QCD scale parameter $\Lambda_\QCD^D=616\,MeV$, for 
the $D$ mesons. 
 The relative error of the data description is $\epsilon^\rho=0.54\%$.

\section{Conclusion}\label{Conclu}
 We have considered bound states as relativistic bound systems in 
the potential approach without using the QCD BS equation or its 
reductions. 
 We have modeled mesons containing light and heavy quarks and their 
resonances in the framework of RQM. 
 We have began our investigation within relativistic classical mechanics 
using the basic principles of symmetries, i.e., the energy and momentum 
conservations' laws in Minkowskiy space. 
 The potential of interaction the Lorentz-scalar function of the spatial 
variable $r$. 
 The concept of position dependent particle mass was used. 
 Using the correspondence principle, we have deduced, from the R2B 
classic equation, the two-particle wave equation. 

 We have calculated masses of light-heavy $S=1$ mesons containing 
$d$ quark and their resonances, i.\,e., $\rho^{\pm}$ and $D^{\pm*}$. 
 Quark masses are close to current masses. 
 We have shown that quarks inside the system move as free particles. 
 Using the complex-mass analysis, we have derived the meson 
interpolating masses formula, in which the real and imaginary parts 
are exact expressions. 
 This approach allows to simultaneously describe in the unified way 
the centered masses of resonances. 
 We have shown here the calculation results for unflavored $\rho$ and 
$D$ mesons and their resonances, however, other mesons, containing 
$s$ and $b$ quarks can be described also well. 

\bibliography{BiDaQM}

\end{document}